\documentclass{aa}
\usepackage{graphicx}

\begin{document}
\title{A new type of photoionized code required 
for the new era of X-ray spectroscopy}

\author{Suzy Collin\inst{1}, Anne-Marie Dumont\inst{1}, Olivier 
Godet\inst{2}}

\offprints{Suzy Collin (suzy.collin@obspm.fr)}

\institute{$^1$LUTH, Observatoire de Paris, Section de
Meudon, F-92195 Meudon Cedex, France\\
$^2$CESR,  9 av. du Colonel Roche, 31028 Toulouse
Cedex 4, France}
\date{Received  : / Accepted  : }

\titlerunning{A new type of photoionized code}
\authorrunning{Suzy Collin et al.}

\abstract{With the advent of the present and future spatial X-ray 
missions, it becomes crucial to model correctly the line spectrum  
of X-ray emitting media such as the photoionized plasma observed  
in the central regions of Active Galactic Nuclei (AGN), or in X-ray 
binaries. 
We have built a photoionization code,  Titan, 
solving the 
transfer of a thousand lines and of the continuum with the ``Accelerated Lambda 
Iteration" method, which is one of the most efficient and at the same 
time the most reliable for line transfer. In all other 
photoionization codes
the line intensities are computed with the 
so-called ``escape probability formalism", used in its simplest 
approximation. In a previous paper 
(Dumont et al. 2003), it was shown that this approximation 
leads to a wrong estimation of the emitted 
X-ray line intensities, especially in the soft X-ray range. The errors 
can exceed one order of 
magnitude in the case of thick media (Thomson thickness of the order 
of unity). In the present paper, we show that
 it also happens, but for different reasons, in the case of
moderately thin media (Thomson thickness of 0.001 to 0.1), 
characteristic of the Warm Absorber in Seyfert 1 or of the X-ray 
emitting medium in Seyfert 2.
Typically, the errors on the line fluxes 
and line ratios 
are of the order of 30$\%$ for a column density of 10$^{20}$ 
cm$^{-2}$, and a factor five for a column density of 10$^{23}$ 
cm$^{-2}$, in conditions giving rise to the spectra observed in 
these objects. 
 We explain why this problem is less acute in 
 cooler media, like the Broad Line Region of AGN. 
  We show some examples of X-ray spectra 
appropriate for Seyfert 2 and for the Warm Absorber of Seyfert 1.  We conclude
that though it is quite important to introduce numerous accurate X-ray data in 
photoionization codes, it should be accompanied by 
 more elaborate methods than escape 
probability approximations to solve the line transfer.

}

\maketitle

\section{Introduction}

Since more than three decades, 
photoionization codes have been developed to compute the structure and 
 the spectrum of photoionized media, such as HII regions, planetary 
 nebulae, supernova remnants, envelopes of novae, Narrow Line Regions of  Active 
Galactic Nuclei (AGN), etc\ldots At the end of the seventies, these codes 
 have been extended to denser and thicker media, like the Broad Emission Line 
 Region (BLR) in quasars and AGN or the emission regions of X-ray 
binaries and cataclysmic variables. For this purpose, the formalism 
of ``escape probability" has been introduced to take into account 
self-absorption in lines while avoiding to solve the 
line transfer (Netzer 1975), and a new type of 
photoionization code has begun to be built.  
 These codes - Cloudy (Ferland et al. 1988), 
XSTAR (Kallman \& Krolik 1995), ION (Netzer 1993), for instance - have now reached 
a high 
degree of sophistication, including very accurate atomic data, a large 
number of ions and transitions, and all the necessary processes allowing to use them 
in various physical conditions. 

With the advent of the X-ray missions Chandra and XMM Newton, splendid spectra 
of various
types of objects have been obtained in the soft X-ray range, showing 
tens of emission lines which can be used as diagnostics of the 
physical state of the emitting region.
The best examples are Seyfert 2 galaxies, which display a rich 
X-ray line
spectrum, 
 most probably 
produced by the external part of the ``Warm Absorber" of Seyfert 1, 
 photoionized by the intense central continuum and seen in 
emission because the central continuum is hidden from 
our view (Antonucci \& Miller 1985).
Typical column densities of this medium are 10$^{21-23}$ cm$^{-2}$ 
(Sako et al. 2000, Kinkhabwala et al. 2002, Ogle et al. 2003).

Most naturally, the photoionized codes developed for the BLR have been 
used to model the X-ray emitting regions of different 
 objects, like the atmosphere of cataclysmic variables, X-ray 
binaries, etc\ldots. These codes are also used to model the absorption lines 
observed in the UV and X-ray range in quasars and AGN. 
 Numerous new X-ray atomic data have been 
introduced in the codes, in order to obtain the best possible accuracy on 
the X-ray spectrum. But the ``escape probability approximation" 
still lies at the center of the computation of the line intensities. 

This formalism, developed in the sixties and in the 
seventies, can be very useful to perform rapid approximate 
computations, but it does not lead to a correct estimation of the source 
function, especially in the case of strongly interlocked transitions, 
including continuum ones. Moreover it uses as a 
local quantity a global one, computed by an integration over 
the whole medium; this is a dramatic extrapolation for very inhomogeneous 
media like
photoionized plasma. These aspects have 
been completely overlooked during the last twenty years, but they become now 
crucial in the context of X-ray emitting media.

In a previous paper, Dumont et al. (2003, referred as D03)  have 
shown that 
 escape probability approximations, at least as they are used in the 
 present codes, are 
unable to compute correct line intensities, within factors of ten, 
when the Thomson thickness of the medium is of the order of a few units, 
typical for the irradiated atmospheres of accretion discs in AGN 
and X-ray binaries. 
In the 
present paper, we extend this study to less optically thick media, and 
we show that the intensities of emission lines are also not accurately computed for 
parameters typical of the Warm Absorber of Seyfert 1 galaxies 
and of the region giving rise to the 
X-ray spectrum in Seyfert 2 galaxies \footnote{One should note 
that modeling an absorption spectrum is much 
easier than an emission one. It requires only 
a correct computation of the thermal and ionization equilibrium, 
which give the fractional ion abundances, i.e. the populations of the 
ground levels, and thus the equivalent widths of the resonance lines, 
which can be compared to those deduced from the observations through a
 curve of growth analysis implying no line transfer. However it can happen 
(and this is indeed the case for 
the Warm Absorber), that emission lines are also produced by the 
absorbing medium, and one should thus worry about the line transfer.}.  

In the next section, we recall very briefly the essence of the 
problem, and we give a few examples in Section 3.

\section{Real transfer versus escape probability approximation} 

\subsection{Failures of the escape probability approximation}

The escape probability method consists in decoupling
 the statistical equations of the levels giving rise to a line photon,
from the transfer 
equation. This is performed by using in both equations a frequency integrated line 
profile which is 
identified with
 the probability of the photon to 
escape  in a single flight from the medium. The problem has been 
amply treated since the sixties
 (cf. for instance ``Radiation transport in spectral lines", by Athay 
 1972, and several reviews 
 in ``Methods of Radiative Transfer", Kalkoven 1984, ``Numerical 
 Radiative Transfer", Kalkoven 1987, Hubeny 2001\ldots). There are different ways to 
 use these escape probabilities, some being actually as sophisticated and 
 time consuming as real transfer. 
 
 In photoionization codes, the emphasis is put on the atomic physics, 
 to account for as many spectral features as possible, 
 and to get a correct thermal and ionization equilibrium. The escape 
 probability is treated in a ``first order approximation", which amounts to 
 replace in the population equations the net radiative rate of excitation 
between two levels (called the Net Radiative Bracket, or NRB)
 by $n_u A_{\rm ul}\rho _{\rm ul}$, where 
\begin{equation}
\rho _{\rm ul}={\left\{{ n_{\rm u}( A_{\rm ul} + B_{\rm ul}\int J_{\nu}{\psi}_{\nu}d\nu) - 
n_l 
(B_{\rm lu}\int J_{\nu}{\phi}_{\nu} d\nu )}\right\} \over n_u A_{\rm ul}} 
\label{eq-div-flux} 
\end{equation}
is identified with the escape probability ($A_{\rm ul}$, $B_{\rm ul}$ and $B_{\rm lu}$ are the 
 Einstein excitation and deexcitation 
coefficients between the upper (u) and lower (l) levels of the 
transition, and $n_{\rm u}$ and $n_{\rm l}$ are the number densities 
of the upper and lower levels). 
The escape 
probability intervenes also in
 the ionization and 
thermal equilibria, through the reabsorption of line photons, 
and in the computation of
the line fluxes emerging from both sides. 

 The escape probability is a {\it global} quantity, expressed as a 
 function of the 
optical thickness at the line center, $\tau_0$, between the 
emission point and the surface, but it depends through the damping 
constant (cf. the appendix) also on physical parameters 
which vary in the medium, such as the temperature and the density. The local 
conditions are thus assumed to be valid all along the photon path. 
 The escape probability is 
 computed by an integration over the whole medium, 
 considered as being homogeneous, which is far from being the case in 
 photoionized plasma except when they are very thin. On the other hand, 
 this global quantity is 
used in the determination of {\it local} rates, such as 
ionizations by line photons.

 The problem is complicated when 
destruction processes occur before the escape of the photon 
(on the spot or along the photon path). This is the 
case if the medium is optically thick in the continuum
 underlying the 
line photon. Various approximate formulae are used to take continuum 
absorption
 into account in the escape probabilities. 
 Note that in hot media, 
 Comptonization acts as an absorption-diffusion process which 
 should also be taken into account in the line transfer.
 
 As shown in D03, no escape approximation
 can give accurate results when the Thomson thickness is of the order 
 of unity or larger. Indeed local reabsorption of line photons by continuum 
processes implies a delicate balance between excitations of X-ray 
transitions by the intense underlying
diffuse X-ray continuum and the net rate
of excitations by the diffuse line flux.  It is not taken properly into account
 in escape 
probability approximations and it creates large differences in the emerging 
line
spectrum. 

When the medium  is only moderately thick (Thomson thickness from 0.001 to 
0.1), these processes are less 
important, but still the escape probabilities lead to inexact results, 
then
for other reasons. For optically thick 
transitions,
escape probabilities
 describe correctly the behavior of the source function deeply in 
 the medium but not at the surface. These lines  are formed 
  close to the surface 
 where their upper level populations are not computed correctly. This 
can be dramatic if the transition is strongly interlocked with other levels 
in a ``multi-level" description. This is the case for the high
 resonance lines and the Balmer lines in hydrogen-like ions,
  or for the 
 resonance lines and the intercombination and forbidden lines for 
 helium-like ions. Avrett and 
Loeser (1987) have shown that, 
 unless methods coupling transfer 
 and escape probabilities are used, large errors follow on the 
 relative  intensities of the lines even for a three-level atom. 
  
 Another problem rises for moderately thick or thin media. Whereas in a thick medium, 
the incident continuum is absorbed completely in resonance 
transitions and it does not play any 
role in line excitation (at least for the most intense lines),
 in the case of thin or moderately thick media as those considered in the 
present paper, the incident continuum can contribute appreciably
 to the line excitation. Indeed it is
 absorbed only at the line center, and a   
fraction is left in the
wings. Its contribution should be 
added to the population equations, as  
 $\rho _{\rm ul}$ includes only the {\it 
diffuse} radiation, which in this case is weak and does not contribute 
much
to the excitations. The
 attenuation factor of the incident continuum is of the order of 
twice the escape 
probability towards the illuminated side (as it is defined in the 
appendix, cf. Elitzur 1982), but it is again an 
approximation. When the medium is 
extremely thin, this factor is equal to unity, and the computation is 
exact.

Moreover, there is also the problem 
of frequency redistribution inside the lines.
Resonance lines are broadened mainly by 
radiation damping, so the emission process is coherent in the frame of 
the atom. It is well-known since a long time that in this case the assumption of 
complete redistribution in a 
Voigt profile leads to strongly overestimate the line intensity (cf. 
for instance Milkey 
\& Mihalas 1973, and Vernazza et al. 1981, who studied L$\alpha$  in 
the solar spectrum), and a way to account better for partial redistribution (PRD)
 is to assume complete redistribution in the Doppler core.  XSTAR
assumes 
complete redistribution in a Doppler core for all lines, whereas in Cloudy, 
  strong resonance lines are treated
with partial redistribution within a Voigt profile, which amounts using 
for the escape probability an expression computed by Hummer \& Kunasz 
(1980) $P_{\rm esc}(\tau_0) = [1+ b(\tau_0)\tau_0]^{-1}$, where $b$ 
of the order of a few units. Note that this function is given only for a 
 ratio of the continuum to line opacity smaller than 10$^{-6}$, and it 
 should be extrapolated for larger values, a 
common case in the present computations. Both 
treatments are close  
to complete redistribution in the Doppler core, and
 similar to the one used here for the resonance 
lines in our escape treatment (cf. the appendix).
 
In summary, the escape probability method is strongly approximated, both 
in very thick and in moderately thick media, for different 
reasons. 

One may therefore ask why it is possible to use this 
method to compute the optical-UV spectrum of the BLR, and 
not  the
X-ray spectrum of a photoionized medium.  

Consider first an X-ray emitting medium photoionized by an intense X-ray continuum, 
i.e. with a large ``ionization parameter", defined as the 
radiation flux to the gas density ratio. The heavy elements are highly 
ionized.
  If the medium is only moderately thick,
the continuum underlying the X-ray lines can nevertheless be optically thick, owing 
 to the presence of several ion species in important proportions at 
 a given point in the medium  (we will illustrate this discussion 
 later in showing which ions are 
 important for absorbing the continuum at the position of the OVIII 
 L$\alpha$ line). 
 
 \begin{table*}
\begin{center}
$\begin{array}{llllllll}
 \multicolumn{4}{c}{\rm Model} & \multicolumn{2}{c}{\tau\rm(cont)\, at\, O\,VIII\,L\alpha} & \multicolumn{2}{c}{\tau_0\rm(O\,VIII\,L\alpha)} \\

\hline   
 \rm No & \rm CD (cm^{-2}) & \xi & \rm V_{turb} (km.s^{-1})& \rm ALI & \rm escape & \rm ALI &\rm escape \\
\hline
\hline 
 & & & & & &  &\\
1 & 10^{15} & 10 & 0 & 8\,10^{-8} &  8\,10^{-8} & 5.3\, 10^{-4} & 5.3\, 10^{-4}\\
2 & 10^{18} & 10 & 0 & 8.5\,10^{-5} &  8.5\,10^{-5} & 0.53 &  0.53 \\
3 & 10^{18} & 10 & 0 & 8.7\,10^{-3} &  8.7\,10^{-3} & 53.1 &  53.1\\
4 & 10^{21} & 10 & 0 & 8.7\,10^{-2} &  8.7\,10^{-2} & 506 &  506\\
5 & 10^{22} & 10 & 0 & 1.10 &  1.17 & 2760 &  2680\\
6 & 10^{22} & 10 & 300 & 1.23 &  1.43 & 61 &  32\\
7 & 10^{22} & 100 & 0 & 7.9\,10^{-2} &  9.4\,10^{-2} & 2241 &  1500\\
8 & 10^{22} & 300 & 0 & 9.5\,10^{-3} &   1.05\,10^{-2} & 260 &  178\\
9 & 10^{23} & 100 & 0 & 1.71 &   2.35 & 33180 &  19300\\
10 & 10^{24} & 1000 & 0 & 0.28 &  0.36 & 5328 &  4250\\
\hline 
\end{array}$
\caption{Characteristics of the models displayed in the figures.}
\label{table-modeles}
\end{center}
\end{table*}
 
 Consider now a less ionized medium like the BLR.
The ionization parameter is smaller, the  
 medium is colder, and only infrared, optical and UV 
 lines are 
 emitted. The column density of the ionized region emitting these lines
  depends on the ionization parameter, but it never reaches
 a Thomson thickness of unity. The continuum underlying the lines is optically 
 thin. For instance ultraviolet lines can only be reabsorbed in the 
 Balmer and Paschen continuum, which are optically thin because the 
 excited levels of HI are not populated at low temperature.  
Once emitted,  these line photons are therefore not reabsorbed by continuum 
processes (except possibly by internal 
dust). The situation is thus simpler than for an X-ray emitting 
medium. On the other hand, $\tau_{0}$ is larger in the optical range than in the 
X-ray range by a factor $\nu_{\rm X}/\nu_{\rm opt}$, so resonance lines 
are very thick in a typical BLR cloud, and escape probabilities can 
be used. Finally the influence of the 
attenuated
incident continuum on line excitation is not important, except 
possibly for 
subordinate lines whose lower level is not much populated. It was 
indeed
stressed in Collin-Souffrin \& Dumont (1986) that the use of escape 
probability can lead to wrong results for the computed Balmer intensities.  

To summarize, the use of escape probability for the BLR is not too bad, both 
because the emission lines are either optically very thick (resonance lines) 
or very thin (forbidden lines), and because the continuum underlying 
these lines 
is optically thin.

\begin{figure*}
 \centering
 \includegraphics[width=18cm]{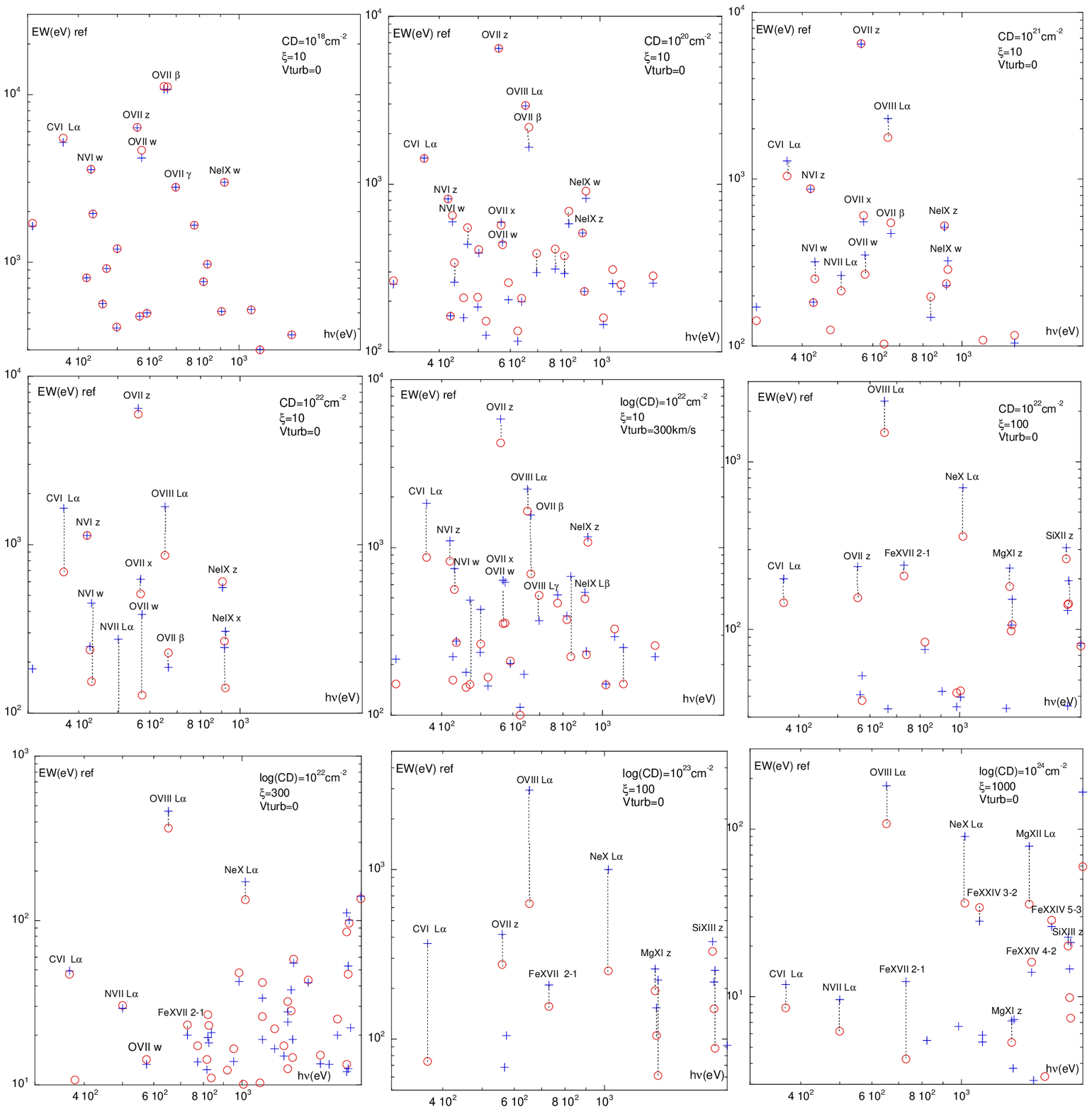} 
\caption{EWs (in eV) of the lines in the 
reflected spectrum, for the different models,
 with the transfer (red circles) and with
the escape treatment (blue crosses). }
\label{fig-Wev}
\end{figure*}

\subsection{Our method}

In order to study X-ray emitting media, we have developed since 
several years a photoionization code, Titan, which 
does not make use of escape probability approximations,
 but solves the transfer both in the lines 
and in the continuum (cf. Dumont et al. 
2000 and Dumont \& Collin 2001). In its last version, Titan 
uses a method called ``Accelerated Lambda Iteration" 
 (ALI), which ensures that all lines 
are computed with an accuracy better than 1$\%$ (cf. D03 for a 
summary of the ALI method). 

Without giving a full description of Titan, it is worthwhile mentioning 
that it includes the 10 most abundant elements (H, He, 
C, N, O, Ne, Mg, Si, S, Fe) and their ion species, i.e. 102 
ions. H-like, He-like, Li-like ions, O~IV and O~V, include a multi-level 
description up to a quantum number $n_{\rm lim}$: $n_{\rm lim}=5$ for H-like, 
 and Li-like ions, and $n_{\rm lim}=3$ He-like ions and for O IV and O V. 
 The other ions 
are treated as two-level atoms plus a continuum (they include several 
resonance lines, but without interlocking). 
 For Li-like ions, we assume 
mixing of the 
different terms of level 3, 
4, and 5. In the case of He-like ions, the atom is made up of all terms 
for $n=2$ and n=3,
and 2 super-levels gather the singlet and triplet levels for n=4.
 Thus the atom is 
represented by 15 levels plus a continuum.  
This better representation is chosen in order to get correct ``triplet 
line" intensities, as they can be separated and are often used as diagnostics for 
the physical conditions of 
X-ray emitting plasma. 
Recombinations onto levels $n > n_{\rm lim}$ are taken into 
account in a simplified way (they are either distributed on all the 
levels, or only on the highest one, whose population is thus 
overestimated). We are thus aware that from the point of view
of atomic data Titan
is not comparable to codes which take 
into account a larger number of 
levels for each ion species, and several thousands of transitions.

Titan solves the ionization equilibrium of all ion species of 
each element, the thermal equilibrium, the statistical equilibrium of 
all levels of each ion - all the physical processes from each 
level being taken into account - coupled with the transfer of the continuum 
and of
about 900 
lines. Comptonization of the continuum above 20 keV is 
taken into account through the coupling with a Monte Carlo code, and 
comptonization
of all lines is computed in an approximate way (cf. D03).  Note 
that though it always plays a role, it does not change appreciably 
the whole line intensity in the 
cases considered here. Moreover, it is treated exactly in the same way 
with the escape and the transfer treatments. 

Finally we should mention that presently Titan is dealing only with 
complete frequency redistribution. In the future we shall 
implement partial redistribution frequency in Titan. Here 
all computations have been performed 
assuming complete redistribution within Doppler core for 
the first resonance line of He- and H-like species, and 
within Voigt profiles for the 
other lines. The latter assumption does not have a strong influence 
on the results, 
since subordinate lines are optically thin or moderately thick, and 
the other resonance lines are not intense. 

\section{Results}

\subsection{The models}

Like in D03, we have compared the spectra obtained when using Titan in its complete 
version,
with those obtained with Titan when the line transfer is replaced by the escape 
probabilities in the computation of the statistical and ionization equilibrium, 
of the energy 
balance, and of the emitted line fluxes. The rest of the computations 
are strictly identical. This ensures that we compare 
exactly the same models. 
Indeed it would not be possible to 
compare directly the results obtained using another photoionization code with those 
of Titan, owing to the difference in the transfer of the continuum 
and in the atomic data. 

In the following computations, we have chosen one of the escape 
probability approximations used in D03, Escape 14 bis, which we 
think to be the best for reasons explained in D03. Radiative excitation 
by the attenuated incident radiation was added, as 
it is important for optically thin media. The equations corresponding 
to this approximation are given in the appendix. 

Our models are plane-parallel slabs of constant hydrogen density $n_{\rm H}=10^{7}$ 
 cm$^{-3}$, and total hydrogen
 column density from $10^{15}$ to $10^{24}$
 cm$^{-2}$, irradiated on one side 
by an incident continuum with a spectral distribution
 $F_{\nu}\propto \nu^{-1}$  extending from 0.1 eV to 100 keV. We have 
 chosen this density, as it is probably intermediate between that of the Warm 
 Absorber in Seyfert 1 and that of the more dilute X-ray emitting medium in Seyfert 2. 
 Moreover, it was shown in Coup\'e et al. (2003) that the density has 
 not a strong impact on the line spectrum between $10^{7}$ and 
 $10^{12}$  
 cm$^{-3}$. It is due to the fact that the most intense lines - resonance 
 lines of H- and He-like ions - are formed by
 recombination. 
We call ionization 
parameter at the surface of the irradiated slab
$\xi = 4 \pi\ F/ n_{\rm H}$,
where $F$ is the integrated incident flux. 
We choose the value of the ionization parameter so as to get spectra dominated in the 
soft X-ray range by H-like 
and He-like lines of C, N, and O, since those are the lines observed
 in Seyfert 1 and 2 spectra. 
We were obliged to increase the value of the ionization 
parameter for the largest column densities, otherwise only a small 
fraction of the slab would be hot and X-ray emitting, and the 
spectra would be 
quite similar to low column density cases. When the column 
density and the ionization parameter are large, several ion species 
are present in successive layers, and the spectrum contains also H- 
and He-like lines of more heavier species (cf.
Coup\'e et al. 2003), so basically the most intense lines 
are shifted towards higher frequencies.  We assumed cosmic abundances 
with respect to hydrogen (Allen 1973): He: 0.085; C: 3.3 10$^{-4}$; 
N: 9.1 10$^{-5}$; O: 6.6 10$^{-4}$; Ne: 8.3 10$^{-5}$; Mg: 2.5 
10$^{-5}$; Si: 3.3 10$^{-5}$; S: 1.6 10$^{-4}$; Fe: 3.2 10$^{-5}$.
 
\begin{figure*}
 \centering
 \includegraphics[width=18cm]{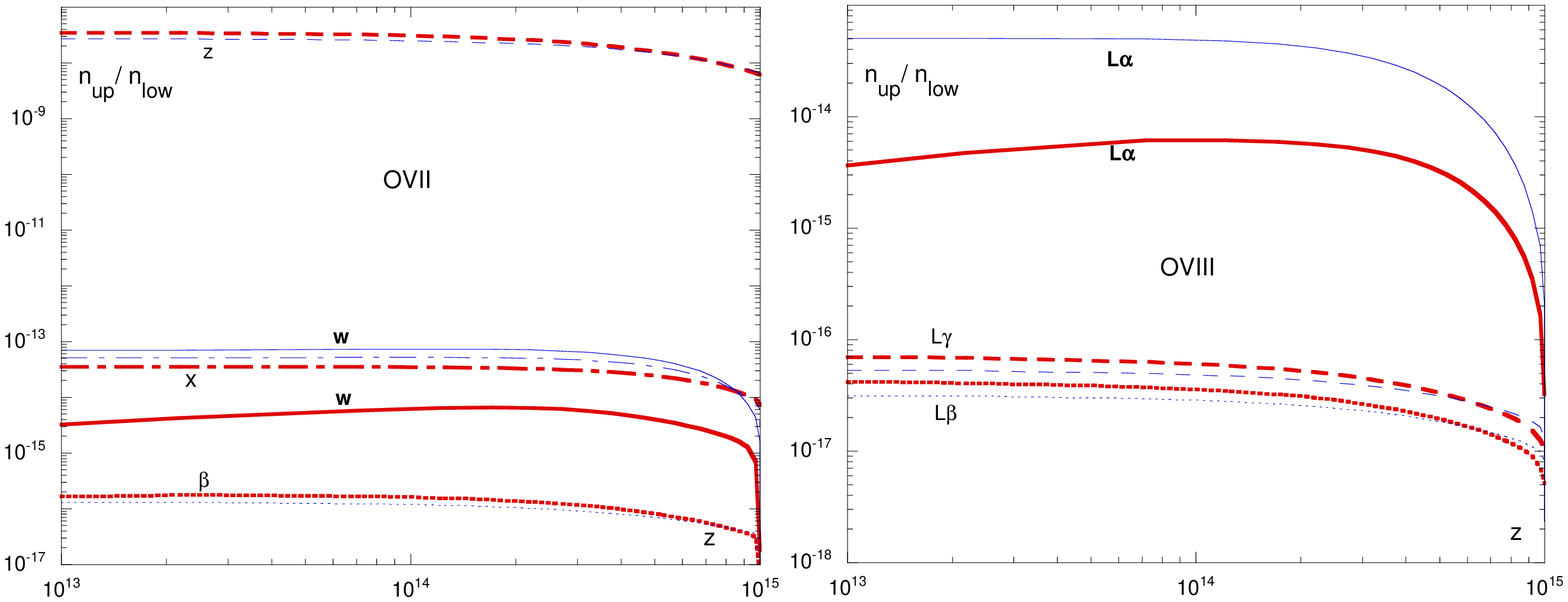}
\caption{Ratios $n_{\rm up}/n_{\rm low}$ as functions of the position in 
 the slab, for several transitions of 
OVIII and OVII, for Model 5 (column density $10^{22}$ cm$^{-2}$, 
$\xi=10$, no turbulent velocity): thick red lines: transfer, thin blue lines: 
escape. }
\label{fig-RapN-c22x10}
\end{figure*}

\begin{figure*}
 \centering
 \includegraphics[width=18cm]{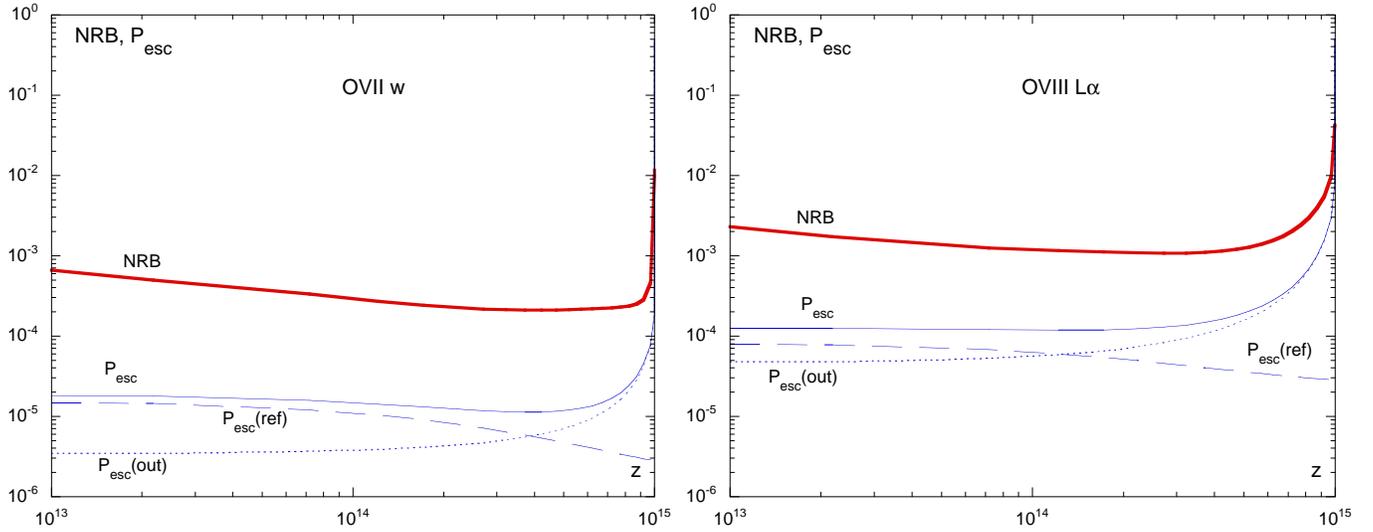}
\caption{NRB (for the transfer) and escape probability $P_{\rm esc}$ (for the escape 
treatment), as functions of the position in 
 the slab, for  
OVIII L$\alpha$ and OVII w, for Model 5 (column density $10^{22}$ cm$^{-2}$, 
$\xi=10$, no turbulent velocity). The figure shows also the escape probabilities 
towards the illuminated and towards the back side, respt. $P_{\rm esc}$(ref) and 
$P_{\rm esc}$(out) (see the 
text for explanations). Note that $P_{\rm esc}$(ref) and 
$P_{\rm esc}$(out) are much smaller than their values at the 
surface, 0.5.  }
\label{fig-c22x10-NRB}
\end{figure*}

\begin{figure}
 \centering
 \includegraphics[width=8cm]{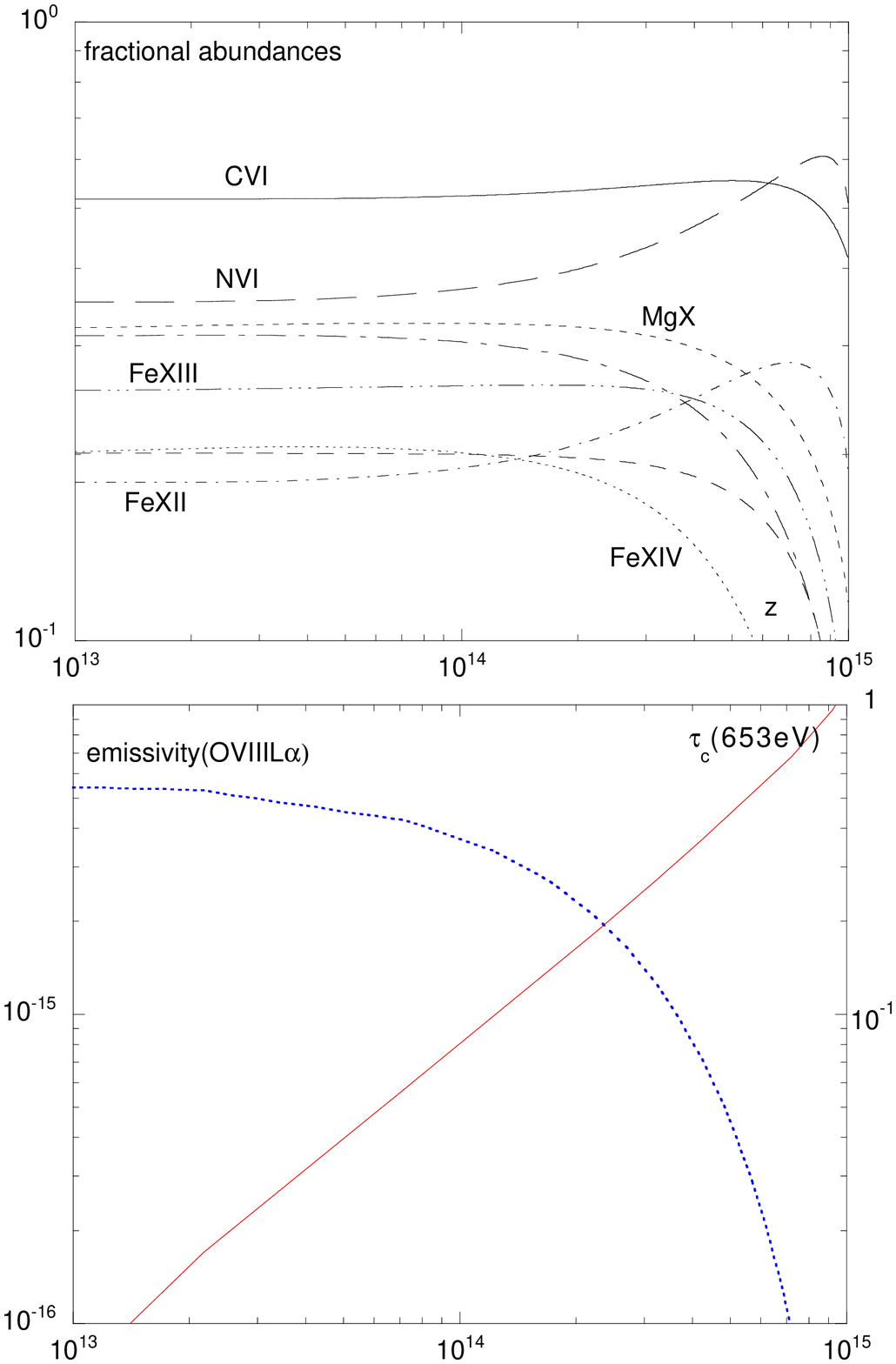}
\caption{ Fractional abundances of several ions able to absorb the
OVIII L$\alpha$ photons (top panel), and
emissivity of the OVIII L$\alpha$ line and optical thickness in the 
continuum at the position of the line (bottom panel),
 as functions of the position in 
 the slab, for Model 5 (column density $10^{22}$ cm$^{-2}$, 
$\xi=10$, no turbulent velocity).}
\label{fig-tauc-LaOVIII}
\end{figure}

In the Warm Absorber or in the emitting medium of Seyfert 
2, a velocity gradient may be present, 
which can decrease
 the optical thickness of the lines.
In the line transfer, it plays the role of
 a micro-turbulent velocity which should be added to the thermal velocity. 
Note that if the emitting medium is made of high velocity clouds 
like the BLR, the velocity gradient acts only as a 
macroscopic velocity which does not influence the transfer  (this 
problem is discussed in more details in Godet et al. 2004).
We have thus run a model with a turbulent velocity of 
300 km/s, close to the measured width of the lines in NGC 1068 (Kinkhabwala et al. 
2002).

Table \ref{table-modeles} summarizes the 
   characteristics of a few models - successively the column density 
   $CD$, the ionization parameter $\xi$, the turbulent velocity 
   $V_{\rm turb}$, the total optical thickness 
   of the continuum at the position of the OVIII  L$\alpha$ line at 
   653eV, and 
   the total optical thickness at the center of the line, both for
    ALI and the 
   escape computation
   \footnote{Since Titan is made for Thomson-thick media, 
   the transfer is treated in the semi-isotropic two-stream Eddington 
   approximation. It means that the optical thickness is multiplied 
   by a factor $\sqrt{3}$ with respect to the normal direction.}.
   
    An immediate result is that the opacity of the 
    continuum increases with the column density up to a limit
    of $\sim 2$. This is because the heavy elements become 
    completely ionized for large values of the ionization 
    parameter, which are required to create a thick hot layer. 
    The same 
    phenomenon exists for $\tau_0$: it cannot exceed 
   a value corresponding to the maximum column density of the 
    ion species, reached for an optimum value of the ionization 
   parameter (cf. Coup\'e et al. 2003 for more detailed 
   explanations). Model 7 with a column density of 10$^{22}$ cm$^{-2}$
   and $\xi=100$ has a continuum opacity 20 times smaller than for 
   $\xi=10$, but it has the same value of $\tau_{0}$(OVIII L$\alpha$),
    because OVIII is still the dominant ion,
     while C and N are already completely ionized.  On the 
     contrary, Model 8 with 
     the same column density but
  $\xi=300$ has a smaller value of $\tau_{0}$(OVIII L$\alpha$),
    because OVIII is no more the dominant ion, it is replaced by OIX. 
     One sees the strong decrease of the line 
   optical thickness in the presence of a large micro-turbulent velocity, 
   while the continuum opacity is almost unchanged. Note also that 
   the continuum opacity and $\tau_{0}$(OVIII L$\alpha$) vary in 
   opposite directions: the first is smaller for the transfer 
   treatment, and the second for the escape approximation. It is 
   linked to the slight variation of the ionization state in the two 
   treatments.

   \subsection{The case of Seyfert 2 : the reflected spectrum}

     We try first to mimic the conditions of the X-ray emitting regions of 
Seyfert 2 galaxies.  According to the Unified Scheme of 
Seyfert galaxies, we are observing in the 
X-ray range the ``reflection" of the central source produced by a 
photoionized 
``mirror", and we do not see the primary photoionizing continuum.
 It is actually not a purely reflecting medium, as it 
reprocesses the radiation (however we shall call this emission the ``reflected" 
spectrum).  
This 
region has probably a conical geometry, and we should observe almost 
perpendicularly to the cone axis. 
 Of course our representation of such a situation can be only very 
 crude, as Titan is dealing with a plane-parallel  
and not a spherical or conical geometry.

Since lines having the largest equivalent widths (EWs) are best observed, 
and since very often EWs are given in the literature,
  we display the results as EWs with respect to the reflected continuum 
  instead of line fluxes.
   Note that the
   spectrum ``reflected" by the illuminated side is 
  similar to that emitted by the back side only in the case of very thin 
  slabs, but it is different for moderately thick slabs. 

 Fig. 
\ref{fig-Wev} displays the EWs of the lines in the 
reflected spectrum, 
for the chosen models and for the transfer and 
the escape treatments. A few intense lines are identified: the spectra 
are dominated by L$\alpha$ lines of H-like and He-like 
lines\footnote{In He-like ions, the first resonant line ($1s^2\ ^1S\ - 2p\ ^1P^o$) 
is called w, the 
forbidden line $1s^2\ ^1S\ - 2s\ ^3S$ is called z, 
and the intercombination 
lines $1s^2\ ^1S\ - 2p\ ^3P^o_{1,2}$ are called x and y.}.
 For larger values of the ionization parameter,  
heavier H- and He-like ions are present. 

\begin{figure}
 \centering
 \includegraphics[width=9cm]{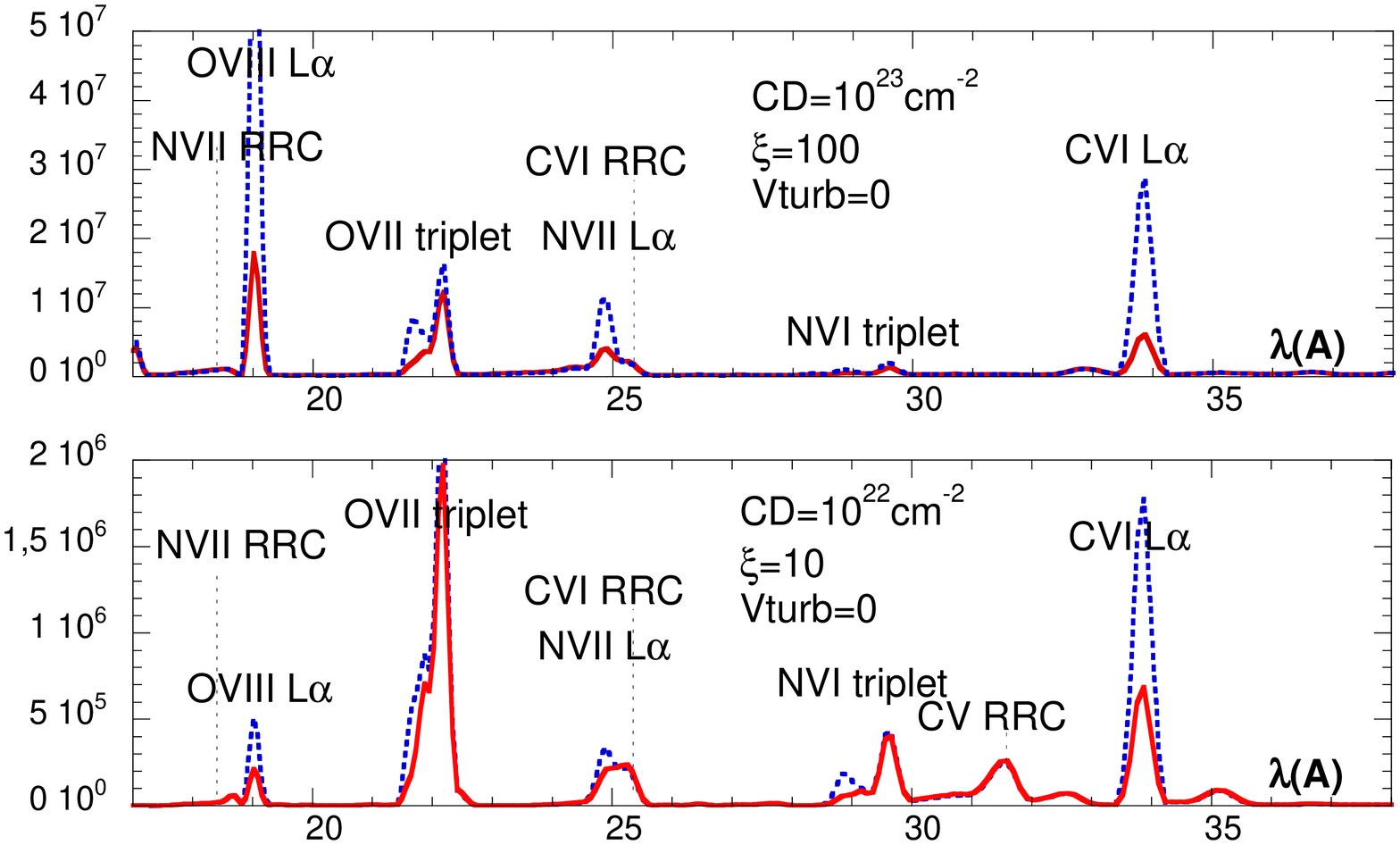}
\caption{ Reflected spectrum for two models: solid red line: transfer; 
dashed blue line: escape probability. The ordinate is proportional to 
$F_{\lambda}$ in photon number
(actually it is equal to
$\nu F_{\nu}$). The spectra are displayed with a spectral resolution of R=30.} 
\label{fig-layout-spe}
\end{figure}
 
The
transfer and the escape treatments give different results
 already at very low column 
densities: for Model 3 (10$^{20}$ cm$^{-2}$), the error is
 30$\%$ for a few lines. It is clearly not due to the influence of 
 the continuum opacity, which is negligible, but to the 
 interlocking between
levels (and perhaps also partly to the approximate treatment of the incident 
 continuum, as $\tau_{0}$ 
 is not negligible). 
 A noticeable result is that, except for very 
 low values of the column density, {\it the escape treatment leads to a 
 systematic 
 overestimation of the resonance line intensities}, as already noticed in D03. 
 It is particularly obvious for 
 L$\alpha$ of H-like ions, and in a lesser amount for the w-term of 
 He-like ions. The overestimation is smaller in the case of forbidden 
 or subordinate lines. 
 
 Figs. \ref{fig-RapN-c22x10} and \ref{fig-c22x10-NRB}  
 can help to understand these discrepancies. Fig.~\ref{fig-RapN-c22x10} displays the 
 ratios of the upper to 
 lower populations $N_{\rm up}/N_{\rm low}$,
 as functions of the position in 
 the slab, for several transitions of 
OVIII and OVII, for Model 5 (column density $10^{22}$ cm$^{-2}$, 
$\xi=10$, no turbulent velocity). This ratio is proportional to the 
source function of the lines. The figure shows that for 
the two resonance lines, OVIII L$\alpha$ and OVII w, it differs strongly
 in the whole slab with the two treatments,
  contrary to those of the other transitions. 
Fig. \ref{fig-c22x10-NRB} displays the NRB (for the transfer) and the escape
 probability $P_{\rm esc}$ (for the escape 
treatment), as functions of the position in 
 the slab, for  
these two lines. $P_{\rm esc}$ is the sum of 
the escape probabilities 
towards the illuminated and towards the back side
 (cf. the appendix), which are also shown on the 
 figure. In principle it should be equal to the NRB, but we see that 
 they have almost nothing in common.
The reflected flux is mainly provided by the region located at 
$10^{13} \le z \le 10^{14}$ cm from the surface, where the source 
functions are overestimated
 by more than one order of magnitude with the escape treatment.

 To also help understanding the difference with a ``cold" medium 
like the BLR, we show for the same model the fractional 
abundances of several ions able to efficiently absorb the OVIII 
L$\alpha$ photons on the top panel of Fig. \ref {fig-tauc-LaOVIII}.
 We see that they are dominant through the whole slab. 
The bottom panel displays the emissivity of the OVIII 
L$\alpha$ line in the escape 
computation, as a function of the depth, and the optical 
thickness in the continuum at the position of the 
line, starting from the illuminated side. The emissivity is important 
up to $z=5\ 10^{14}$ cm, where the optical thickness reaches a 
value of the order of 0.5. Thus the reflected line is attenuated by 
about a factor two. However it is not sufficient to explain the 
discrepancy between the escape and the transfer treatments, and one 
should also invoke the other reasons mentioned previously. 

\begin{figure}
 \centering
 \includegraphics[width=8cm]{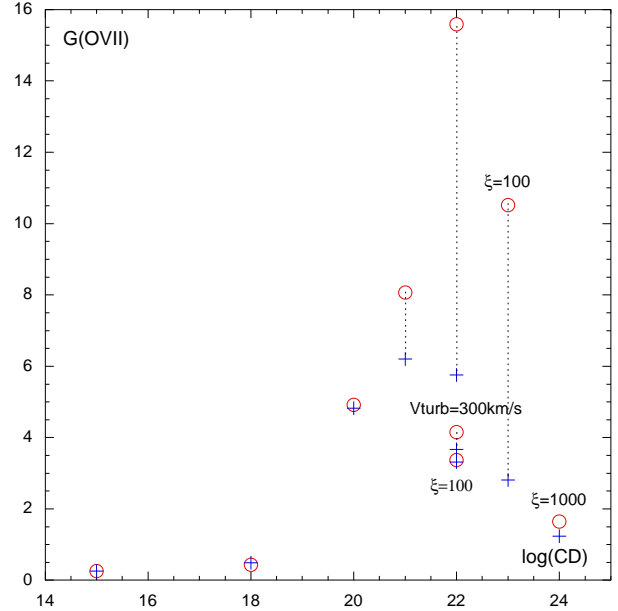}
\caption{G ratios (see the text) in the 
reflected spectrum versus the column density,
 for the transfer (red circles) and 
the escape treatment (blue crosses). 
When another value is not indicated, $\xi=10$. }
\label{fig-G-ech-ali}
\end{figure}

 The errors on the 
 escape results increase with the column density and reach a factor 
 five for Model 9 (10$^{23}$ cm$^{-2}$, $\xi= 100$), where both the 
 line and the continuum opacities are large, so the influence of the 
 diffuse continuum becomes important.They
 decrease for Model 10 (10$^{24}$ cm$^{-2}$, $\xi= 1000$), 
 because the opacities 
 of the continuum and the lines are smaller.  Note also that the errors 
 are quite small with Model 8 (10$^{22}$ cm$^{-2}$, $\xi= 300$). It 
 is easily understood: this model is highly ionized, and the 
 column density is not large enough to allow the presence of low 
 ionized species. As a result, the spectrum is mostly made of He- and 
 H-like lines of heavy elements and of OVIII, and it does not display the OVII
 lines  characteristic of a Seyfert 2 spectrum, as can be seen on 
 Fig. \ref{fig-Wev}.
 
 The comparison of Models 5 and 6
 (10$^{22}$ cm$^{-2}$, $\xi= 10$, respectively without and with 
 a micro-turbulent velocity of 300 km/sec) 
 shows that  
the escape probability approximation gives slightly better results in the 
presence of a micro-turbulent velocity, as expected since the lines 
are less optically thick. Still the 
errors on line intensities and line ratios are typically of the order 
of a factor two. Moreover, 
the results are shifted towards 
smaller column densities for a given EW (Godet et al. 2004): 
here, for a turbulent velocity of 300 km/sec, the EWs of the resonance 
lines are similar to those without turbulent velocity, but for a column density 
about one order of magnitude smaller.  The comparison 
of Model 5 with Model 7 (10$^{22}$ cm$^{-2}$, $\xi= 100$)
 shows also that 
the escape probability approximation gives a better result in the 
latter case, because the continuum opacity is small. 
Also, due to the higher ionization parameter, the spectrum displays more intense 
lines of He- and H-like heavy 
elements. 

In summary, the errors in the line intensities and line ratios 
due to the use of the escape approximation are 
not easy to predict, as they depend on different processes 
in a complex and subtle way. They are important even for low column 
densities.   

\medskip

To illustrate more clearly the effects of the use of escape 
probabilities on the observed spectrum, Figs. 
\ref{fig-layout-spe} displays the reflected spectrum for two models, 
in the same form as they 
are generally published in the literature. Though the column densities 
are relatively small (10$^{22}$  and 10$^{23}$ cm$^{-2}$), large 
discrepancies appear between the escape and the transfer treatments, 
which can lead to misinterpretations of the observed spectra when 
modeled by the escape treatment.

The ratio G=(z+x+y)/w of He-like ions is used as a diagnostic 
for hot plasma. 
The variation of this ratio as a function of several parameters is 
studied in Coup\'e et al. 2003 and Godet et al. 2004. Here we focus only 
on the differences between the escape and the transfer treatment. 
 Fig. \ref{fig-G-ech-ali} displays this ratio versus the column 
 density for the models studied here.
Again we see that the differences are the largest (a factor three) 
for Models 5 and 8. It is 
however interesting to notice that in the 
presence of a turbulent velocity, G is almost the same for both 
treatments.

\begin{figure*}
 \centering
 \includegraphics[width=18cm]{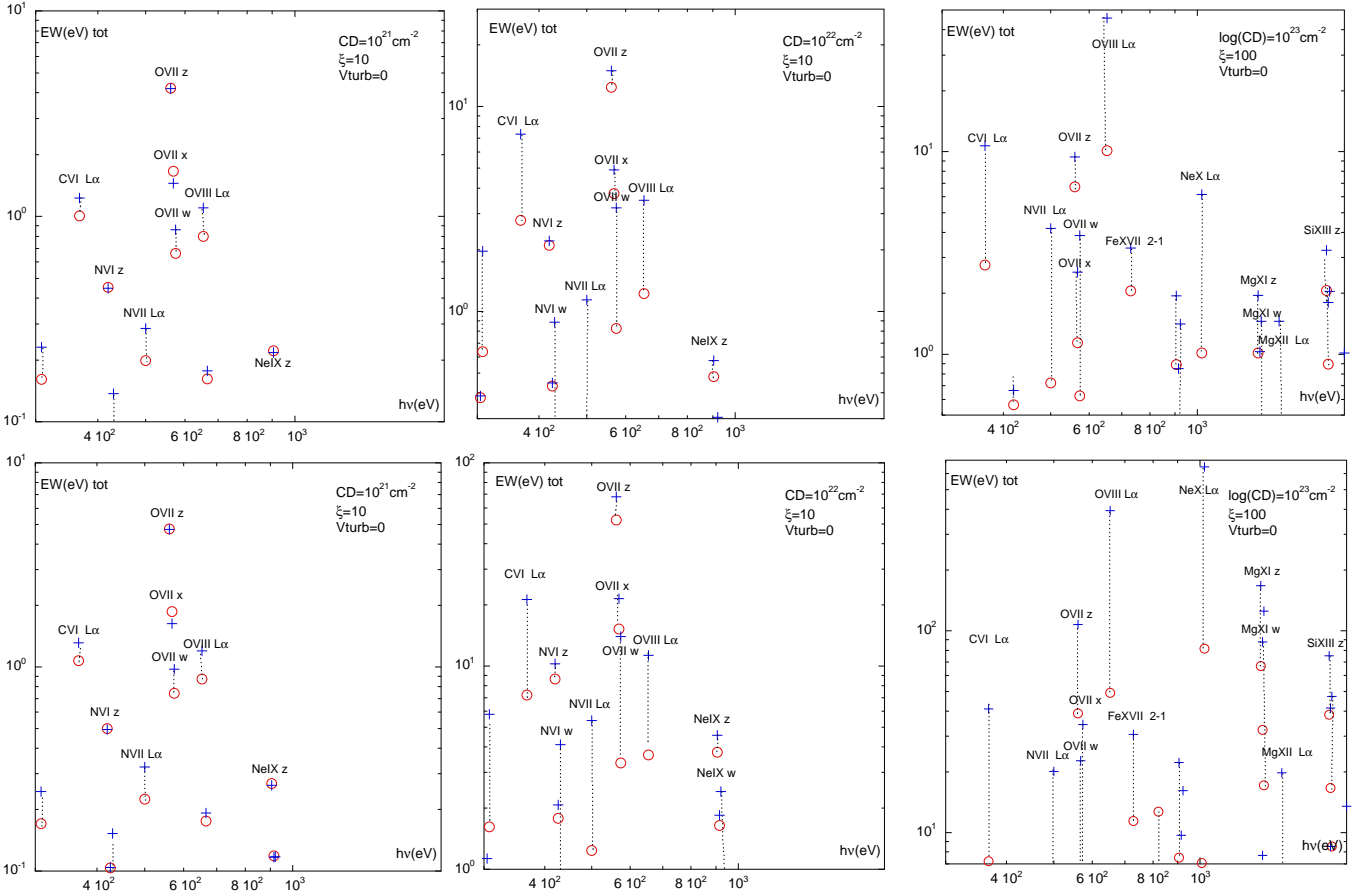}
\caption{ EWs (in eV) of the lines, in the case of a spherical ``Warm 
Absorber" with  a coverage 
factor of 0.5, for different models computed 
 with the transfer (red circles) and with
the escape treatment (blue crosses). The top panels correspond 
to the line of sight of the central X-ray 
source not crossing the Warm Absorber,
 and the bottom panels to the Warm Absorber crossing
 the line of sight (cf. the text).}
\label{fig-Wevtot}
\end{figure*}

\subsection{The case of Seyfert 1: the emitted spectrum}

In Seyfert 1, one observes directly the central X-ray continuum, 
possibly after
absorption by the Warm Absorber. About 50$\%$ 
of Seyfert galaxies have absorption lines, and it is most 
probable that the ``Warm Absorber" is present 
in all active nuclei, but it does not cross always the line of sight 
of the X-ray continuum. Emission lines are produced by 
the Warm 
Absorber, but they are diluted by the presence of the primary continuum.
 The observed
spectrum is thus the sum of reflection from the illuminated side and 
emission from
the back side of the clouds, plus the primary continuum, absorbed 
if the line of sight crosses the Warm Absorber.
If it is distributed spherically or axi-symmetrically, 
the  proportions coming from each side are identical. 
Assuming  a 
coverage factor $f$, the observed EWs are thus: 
\begin{equation}
EW={f(1-f)EW({\rm ref})F_{\nu}({\rm c,ref})+f 
EW({\rm out})F_{\nu}({\rm c,out})\over 
f(1-f)F_{\nu}({\rm c,ref})+f 
F_{\nu}({\rm c,out})+F_{\nu}({\rm c,prim})}
\label{eq-Wevtot}
\end{equation}
 where 
$F_{\nu}({\rm c,ref})$, $F_{\nu}({\rm c,out})$, $F_{\nu}({\rm c,prim})$, are 
the continuum flux, respectively  reflected, emitted outward, and the
primary continuum (absorbed or not), at the line frequency. 
 If the line of sight crosses the Warm Absorber, there are also absorption lines 
 imprinted on the primary continuum. We will assume that they are not 
 located at 
the same frequencies as the emission lines, owing to the existence of 
velocity gradients 
(for instance in the case of a wind, the emission lines are
redshifted
with respect to the absorption lines).

As an illustration, Fig.\ref{fig-Wevtot} displays the EWs of the 
emission lines produced by the Warm Absorber with a covering factor 
$f=0.5$: 1- not crossing the line of sight (top panels), 
2- crossing
the line of sight (bottom panels). Comparing these results to those of 
Seyfert 2, we see that the emission line spectra are very similar in 
both cases. 
However when 
the primary continuum is not absorbed (top panel), the EWs are always 
very small, and only a few lines could be detectable (OVIIz, OVIII 
L$\alpha$, CVI L$\alpha$). When the line of sight crosses the Warm 
Absorber, the EWs are larger, 
and many lines should be detected for a
column density of the order of 10$^{22}$ cm$^{-2}$ or larger. So it is a 
powerful way to estimate the column density of the Warm Absorber, 
independent of the absorption lines. Note also that the line ratios 
differ between the case with and
without absorption of the primary continuum,
 owing to the variation of the continuum 
opacity with the frequency.

Anyway, our purpose in this paper is not to study the emissive regions 
of Seyfert nuclei, but to estimate the validity of the computations 
performed with the escape approximation. Fig.\ref{fig-Wevtot} shows 
that, like in the case of Seyfert 2, the EWs are overestimated by 
large factors (up to 5 for Model 9) by the approximate treatment. 
Again also, the overestimation is much larger for resonance lines than 
for forbidden or subordinate lines.

\section{Conclusion} 

We have shown that the use of the escape probability approximation leads 
to large errors in the computed line fluxes and line ratios, in 
conditions which are typical of the X-ray emitting regions of Seyfert 
2 nuclei, and of the Warm Absorber of Seyfert 1, i.e. for a Thomson  
thickness of the order of 0.001 or larger. This completes the previous 
paper (D03), where the same study was performed for thicker media, 
and where it was shown that the use of the escape approximation leads to 
errors by more than one order of magnitude on the line fluxes and line 
ratios. We find here that the errors are of the order of 30$\%$ for a column 
density of 10$^{20}$ cm$^{-2}$, and can reach a factor five for a 
column density 10$^{23}$ cm$^{-2}$. We confirm that they are 
almost always in the direction of an overestimation of the most intense line intensities, 
especially of the L$\alpha$ line of H-like ions, and of the resonance w 
term of He-like ions. We explain why such large errors 
occur for X-ray emitting media, and not for cooler media like the BLR. 

The comparison between the escape approximation and the transfer 
treatment (performed through the Accelerated Lambda Iteration method
with our photoionization code Titan), is made in such a way that 
no other possible explanation of the discrepancies than the use of the 
escape probability approximation can be invoked. So one is led to conclude that 
unless a real transfer of the lines
is introduced in the codes for modelling X-ray spectra, the results 
cannot have an accuracy better than that given by the approximation, 
even when the treatment of atomic physics is highly sophisticated.

Whatever the discrepancy between the results of the escape 
probability and the transfer treatment, an uncertainty remains 
concerning the real intensity of the resonance lines. Here they were 
computed assuming complete redistribution within Doppler core, which 
mimics partial redistribution within a Voigt profile. PRD 
cannot be taken into account accurately with the escape probability 
formalism, but it can be done with the transfer treatment, provided 
the implementation in the code of
another substantial time consuming procedure. We are presently 
studying such an improvement. 

\bigskip

\noindent {\bf APPENDIX: Equations used for the escape approximation 
in the optically thin case}

We compute the escape probability towards the surface $P_{e}(\tau_0)$ of 
a line  as (cf. Collin-Souffrin et al. 1981):

\begin{equation}
 P_{e}(\tau_0) ={\rm max}(f1,f2),
 \label{eq-esc-1} 
 \end{equation}
for all subordinate and high resonance lines, and
\begin{equation}
 P_{e}(\tau_0) =f1
 \label{eq-esc-2} 
 \end{equation}
for the first resonance lines of H- and He-like species, with
\begin{eqnarray}
 f1 ={0.5\over 1+2\tau_0\sqrt{\pi 
ln(\tau_0+1)}}
&& f2= {2\over
3}\sqrt{{a\over\sqrt{\pi}\tau_0}},
\label{eq-esc-3}
\end{eqnarray}
where $\tau_0$ is the optical thickness at the line center between 
the emission point and the surface (we recall that it is taken along 
the photon path, so it
multiplied by a 
factor $\sqrt{3}$ to take into 
account the fact that the direction of the emitted photon is at 
random, and to be able to compare the escape approximation to the transfer 
treatment), and $a$ is 
the usual damping constant. $f1$ corresponds to
the Doppler core, and $f2$ to the Lorentz wings of the 
Voigt profile.

The total escape probability $P_{\rm esc}$ is the sum of the escape probability 
towards the illuminated side, $\beta_{\rm ref}=P_{\rm e}(\tau_0)$,  and 
towards the back side,
$\beta_{\rm out}=P_{\rm e}(T_0-\tau_0)$, where $T_0$ is the total optical 
thickness of 
the slab
depth at the line center.

In the equations for the level populations, the net radiative rate 
from an excited level is replaced by $A_{\rm ul}\beta_{\rm pop}$, with 
\begin{equation}
\beta_{\rm pop}={\rm min}[1, P_{\rm esc}(\tau) \times (1+{\kappa_{\rm c}\over 
\kappa_l}F({\kappa_{\rm c}\over \sqrt{\kappa_{\rm l}}}))],
\label{eq-betapop}
\end{equation}
where $\kappa_{\rm c}$ and $\kappa_{\rm l}$ are respectively the absorption 
coefficient in the continuum and in the line, and $ F$ is the 
operator given by Hummer (1968) to account for destruction by 
continuum absorption in one line scattering:
\begin{equation}
F(X)=\int_{-\infty}^\infty {\phi (x)\over 
X+\phi (x)} dx
\label{eq-FX}
\end{equation}
where $\phi (x)$ is the absorption line profile,
$x=\delta \nu /\delta \nu _{\rm D}$, and $\delta \nu _{\rm D}$ is the Doppler width.

A term $B_{\rm lu}J_{\nu}^{\rm inc.att.}$ is added to the net radiative rate to 
take into account excitations by the attenuated incident radiation 
$J_{\nu}^{\rm inc.att.}=2\beta_{\rm ref}/(2\pi)F_{\nu}^{\rm  
inc}$, according to the definition of the flux, 
$F_{\nu}^{\rm  inc}$ being the incident flux at the frequency $\nu$ 
(the corresponding deexcitation rate is negligible).  

The local cooling for each line is:
\begin{equation}
\Lambda_{\rm line}=(n_{\rm u}A_{\rm ul} P_{\rm esc}-n_{\rm l}B_{\rm lu}J_{\nu}^{\rm inc.att.})
\ {h\nu\over n_{\rm e}n_{\rm H}} .
\label{eq-lambda-line}
\end{equation}

The reflected flux in a line is computed as:
\begin{equation}
F_{\rm ref}= \int{n_{\rm u}A_{\rm ul}h\nu \beta_{\rm ref}\ 
{\rm exp}[-\tau_{\rm e}] dz},
\label{eq-Fref}
\end{equation}
where $\tau_{\rm e}$ is the effective optical 
thickness  of the slab in the 
continuum at the line frequency,
between the current point and the illuminated surface.

In the outward emitted line flux, one must take into account the 
photons absorbed in the incident continuum:
\begin{eqnarray}
F_{\rm out}&=& \int{(n_{\rm u}A_{\rm ul}h\nu \beta_{\rm out} 
{\rm exp}[-(T_{\rm e}-\tau_{\rm e})]} 
\\
\nonumber
&& \ \ \ -n_{\rm l}B_{\rm lu}J_{\nu}^{\rm inc.att.}) dz,
\label{eq-Fout}
\end{eqnarray} 
where $T_{\rm e}$ is the effective total optical thickness of the slab in the 
continuum at the line frequency. 

The ionization rate 
due to the lines, at the depth $z$, is equal to: 
\begin{eqnarray}
&&\sqrt{3}\kappa_{\rm c} A_{\rm ul}\times 
\\
\nonumber
&& \left( \int_0^z{(n_{\rm u} 
\beta_{\rm out}{\rm exp}[-\tau_{\rm e}+T_{\rm e}]-n_{\rm l}{\rm B_{\rm lu}\over 
 A_{\rm ul}}J_{\nu}^{\rm inc.att.})} 
dZ \right.
\\
\nonumber
 &&\left. + \int_H^z{n_{\rm u} 
\beta_{\rm ref}{\rm exp}[-T_{\rm e}+\tau_{\rm e}]dZ}\right).
\label{eq-ion}
\end{eqnarray} 
This expression is used also for the gains by photoionizations due to the 
lines. 

The escape probability is slightly different in the case of a 
highly ionized and/or moderately thick medium, as one must take into 
account the shift of line photons by comptonization. Thus 
$\beta_{\rm ref}$ is replaced by 
 $\beta'_{\rm ref}= \beta_{\rm ref}+{1\over 2} (1-P_{\rm esc}(\tau)) {\sigma\over 
\kappa_l\sqrt{\pi}+\kappa_c+\sigma}$, and $\beta_{\rm out}$ by
 $\beta'_{\rm out}= \beta_{\rm out}+
{1\over 2} (1-P_{\rm esc}(\tau)) {\sigma\over 
\kappa_l\sqrt{\pi}+\kappa_c+\sigma}$, and $P_{\rm e}$ is replaced by  
$P'_{\rm e}=\beta'_{\rm ref}+\beta'_{\rm out}$.



\bigskip



\end{document}